\documentclass[12pt,preprint]{aastex}

\usepackage{graphicx}

\slugcomment{Submitted to ApJ}
\shorttitle{Photospheric Fluorescent Fe~K$\alpha$}
\shortauthors{Drake et al.}
\begin{document}
\title{X-ray Fluorescent Fe~K$\alpha$ Lines from Stellar Photospheres}
\author{Jeremy J.~Drake,\altaffilmark{1}
Barbara Ercolano,\altaffilmark{1} and 
Douglas A.~Swartz,\altaffilmark{2}}
\affil{$^1$Harvard-Smithsonian Center for Astrophysics,
MS-3, \\ 60 Garden Street, \\ Cambridge, MA 02138}
\affil{$^2$USRA, George C.\ Marshall Space Flight Center, VP62,
  Huntsville, AL 35812} 

\begin{abstract}

X-ray spectra from stellar coronae are reprocessed by the underlying
photosphere through scattering and photoionization events.  While
reprocessed X-ray spectra reaching a distant observer are at a flux
level of only a few percent of that of the corona itself,
characteristic lines formed by inner shell photoionization of some
abundant elements can be significantly stronger.  The emergent
photospheric spectra are sensitive to the distance and location of the
fluorescing radiation and can provide diagnostics of coronal geometry
and abundance.  Here we present Monte Carlo simulations of the
photospheric K$\alpha_1,\alpha_2$ doublet arising from quasi-neutral
Fe irradiated by a coronal X-ray source.  Fluorescent line strengths
have been computed as a function of the height of the radiation
source, the temperature of the ionising X-ray spectrum, and the
viewing angle.  We also illustrate how the fluorescence
efficiencies scale with the photospheric metallicity and the Fe
abundance.  Based on the results we make three comments: (1)
fluorescent Fe lines seen from pre-main sequence stars mostly suggest
flared disk geometries and/or super-solar disk Fe abundances; (2) 
the extreme $\approx 1400$~m\AA\ line observed from a flare on
V~1486~Ori can be explained entirely by X-ray fluorescence if the
flare itself were partially eclipsed by the limb of the star; and (3)
the fluorescent Fe line detected by {\it Swift} during a large flare
on II~Peg is consistent with X-ray excitation and does not require a
collisional ionisation contribution.  There is no convincing evidence
supporting the energetically challenging explanation of electron
impact excitation for observed stellar Fe~K$\alpha$ lines.
\end{abstract}

\section{Introduction}
\label{s:intro}

It is well-established from surveys of the sky at EUV and X-ray
wavelengths that all stars with spectral types later than mid-F,
except for giants later than mid-K, possess hot outer atmospheres akin
to that of the Sun \citep[e.g.][]{Vaiana.etal:81, Schmitt:97}.  While
much observational and theoretical effort has been devoted to
understanding solar coronal spectra and, in more recent years, toward
understanding stellar coronal emission and spectra, comparatively little
attention has been devoted to the reprocessing and line
fluorescence resulting from this coronal emission by the underlying
solar and stellar photospheres.  In contrast, considerable effort has
been spent on the study of X-ray reprocessing by ``cold'' gas in much
more complex systems with more prominent fluorescent features but more
uncertain geometries and physical conditions, such as the accretion
disks around black holes and non-degenerate objects in X-ray binaries
\citep[e.g.][]{Felsteiner.Opher:76,Hatchett.Weaver:77,Fabian.etal:89, 
George.Fabian:91,Laor:91, Matt.etal:97, Ballantyne.etal:02, 
Beckwith.Done:04, Cadez.Calvani:05, Dovciak.etal:04,
Laming.Titarchuk:04, Brenneman.Reynolds:06}. 
The processes involved in photospheric fluorescence by coronal
irradiation are the same as those discussed in these works; the main
difference here is in the specific geometry of the X-ray source above
a quasi-neutral sphere and of the extended, shell-like nature of the 
coronal source above the photosphere during quiescent conditions.

X-rays emitted from a hot ($T\ga 10^6$~K) corona incident on the
underlying photosphere can undergo either Compton scattering or
photoabsorption events through the ionization of atoms or weakly
ionized species. Through scattering events, photons can be reflected
back in a direction towards the stellar surface where they have a
finite chance of escape.  Compton scattering redistributes the
spectrum to lower energies by $\sim E^2/m_ec^2$ per collision, where $E$
is the photon energy and $m_e$ the electron rest mass.  The spectrum
reflected from a stellar surface by scattering events is then shifted
and broadened towards lower energies.  Photoionization events
involving X-ray photons directed toward the photosphere are
predominantly inner shell interactions with astrophysically abundant
elements, the outer and valance cross-sections being very small at
these energies.  Observable fluorescent lines can then arise as a
result of the finite escape probabilities of photons emitted in outward
directions by hole transitions in these atoms photoionized in their
inner shells.  These processes have been described in the solar
context by, e.g., \citet{Tomblin:72} and \citet{Bai:79} (B79), and
more recently for arbitrarily photoionized slabs by \citet{Kallman.etal:04}.

The strongest of the fluorescent lines for a plasma of approximately
cosmic composition is the $2s$-$1p$ 6.4~keV Fe K$\alpha$ doublet
occurring following ejection of a $1s$ electron.  It has been observed
in solar spectra on numerous occasions \citep[e.g.][]{Neupert.etal:67,
Doschek.etal:71, Feldman.etal:80, Tanaka.etal:84,
Parmar.etal:84, Zarro.etal:92}.  The mechanism of fluorescence by the
thermal X-ray coronal continuum was suggested by
\citet{Neupert.etal:67}, and was firmly established on more
theoretical grounds by \citet{Basko:78,Basko:79} and B79.
\citet{Parmar.etal:84} provided convincing observational confirmation
based on flare spectra obtained by the {\it Solar Maximum Mission},
though it has also been noted that contributions from non-thermal
electron impact might also be present during hard X-ray bursts 
\citep*[e.g.][]{Emslie.etal:86, Zarro.etal:92}.

B79 pointed out that, for a given source spectrum, the observed flux
of Fe~K$\alpha$ photons from the photosphere depends on essentially
three parameters: the photospheric iron abundance; the height of the
emitting source; and the heliocentric angle between the emitting
source and observer.  \citet{Phillips.etal:94} used Fe K$\beta$
observations to probe the difference between the photospheric and
coronal iron abundance for flares observed by the {\it Yohkoh}
satellite.  More importantly for the stellar case, the spatial aspects
of photospheric fluorescent line formation suggest its application to
understanding the spatial distribution of coronal structures and
flares on stars of different spectral type and activity level to the
Sun \citep{Drake.etal:99}.  Indeed, Fe~K fluorescence has recently
been detected during flares on the active binary II~Peg
\citep{Osten.etal:07} and on the single giant HR~9024
\citep{Testa.etal:07}.  The Fe~K line has also been seen in a growing
sample of pre-main sequence (PMS) stars
\citep{Imanishi.etal:01,Favata.etal:05,Tsujimoto.etal:05,
Giardino.etal:07}, in which the line is thought to originate
predominantly from the irradiated protoplanetary disk rather than the
photosphere.


Since the work of \cite{Bai:79}, there have been no concerted efforts
to extend models of photospheric fluorescence for coronal excitation
sources with other characteristics.  Fluorescent lines other than
Fe~K$\alpha$ have also not yet, to our knowledge, been studied by
other workers in any detail in this context.  A reasonably strong
feature observed in solar spectra near 17.62~\AA\ had been identified
with Fe L$\alpha$ photospheric fluorescence \citep{McKenzie.etal:80,
Phillips.etal:82}, but calculations of the expected line strength was
shown by \citet{Drake.etal:99} to be much too weak to explain the
feature, and these authors instead identified the line with a
transition in Fe~XVIII arising from configurational mixing and both
seen in {\it Electron Beam Ion Trap} spectra and predicted by theory
\citep{Cornille.etal:92}.  However, given the potential diagnostic
value of photospheric fluorescence, other lines, such as O~K$\alpha$,
are possibly observable with very high quality observations and
warrant further study.

The capabilities of current X-ray missions such as {\it Chandra}, {\it
XMM-Newton}, {\it Swift} and {\it Suzaku} to detect fluorescent lines
further motivates a re-examination of the photospheric fluorescence
problem in the context of stellar coronae, photospheres and
protoplanetary disks.  We restrict the study in hand to Fe~K
fluorescence from stellar photospheres and defer detailed discussions of
protoplanetary disks and fluorescence from other elements to future
work.

\section{Calculations of Photospheric Fluorescent Spectra}
\label{s:calcs}


\subsection{Photospheric Penetration of Coronal X-rays}
\label{s:penetration}

\begin{figure}
\epsscale{1.0}
\plotone{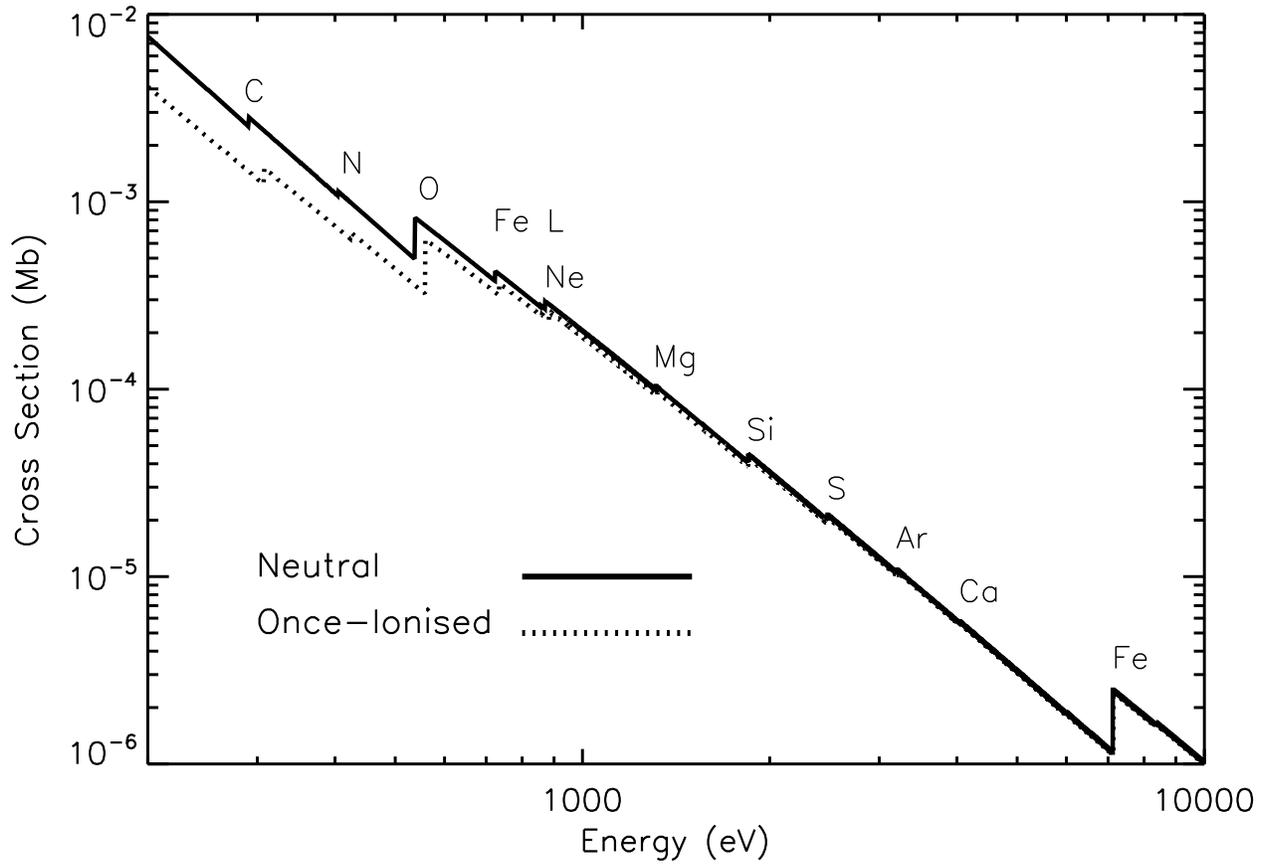}
\caption{\footnotesize 
The photoabsorption cross-section in the energy
range of interest for a neutral gas (solid curve) and once-ionised
(dashed curve) plasma of solar composition.  The labels indicate the
ionisation edges for most abundant elements.
\label{f:cross}
}
\end{figure}

In order to establish the atmospheric conditions under which
fluorescence by, and Compton scattering of, coronal X-rays take place,
it is necessary to know where in a stellar atmosphere incident X-rays
are absorbed.  Since the absorption cross-section for a cool plasma
with solar, or near-solar, composition is a strong function of photon
energy, the absorption of the coronal X-ray spectrum will take place
at different levels in the atmosphere.  To determine the
characteristic height of absorption as a function of photon energy we
have computed the optical depth for X-rays as a function of altitude
in the solar atmosphere.  We adopted the \citet*{Vernazza.etal:81}
Model C (VALC) describing the solar atmospheric
structure as a function of altitude, and used the photoionization
cross-sections from \citet{Verner.etal:93} and \citet{Verner.Yakovlev:95}
to compute the optical depth.  The total absorption cross-section for a
plasma with the solar chemical composition of
\citet{Grevesse.Sauval:98} (GS)
in the energy range of interest for this study is illustrated in
Figure~\ref{f:cross} for the case of both a neutral gas and a plasma
in which all species are once-ionized (this latter situation does not 
correspond to any physically valid thermal equilibrium condition, but 
is simply for comparative purposes).  

\begin{figure}
\epsscale{1.0}
\plotone{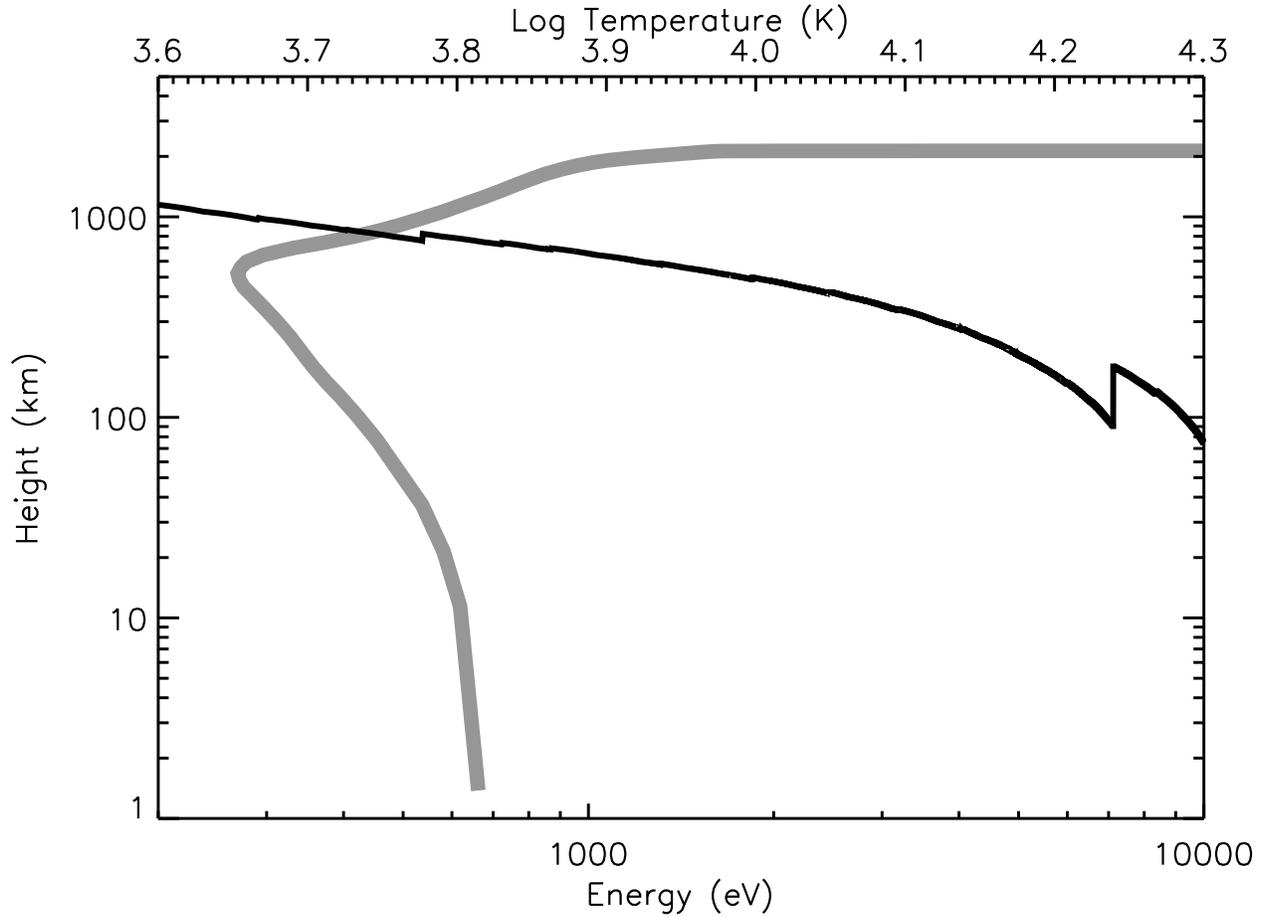}
\caption{\footnotesize
The locus corresponding to optical depth of unity for X-rays incident
on the solar photospheric VALC model plotted as a function of incident
photon energy and height (thin black curve).  Toward lower energies
absorption occurs predominantly in the lower chromosphere; this region
is shifted downwards with increasing energy, reaching the upper
photosphere at the Fe K photoionization threshold energy.  The model
temperature structure as a function of height is illustrated by the
thick grey curve.
\label{f:depth}
}
\end{figure}

The height in the VALC model corresponding to optical depth of unity
for X-rays is illustrated in Figure~\ref{f:depth}.  Also
illustrated is the atmospheric temperature as a function of height in
this figure for comparison purposes.  At the energy corresponding to 
the absorption edge of C, optical
depth of unity is reached in the lower chromosphere, at temperatures
of $\sim 6000$~K and about 1000~km above the point where the continuum
optical depth at 5000 \AA, $\tau_{5000}$, is unity (defined to be at a
height of 0~km in this model).  Towards higher energies, the
decreasing photoabsorption cross-section leads to increasingly deeper
penetration of coronal X-rays.  At the Fe K photoionization threshold
energy, the typical height of absorption is well into the photosphere
and only 100~km or so above $\tau_{5000}=1$.  At this point in the
solar atmosphere, most Fe is in the form of Fe$^+$ and the temperature
is $\sim 5400$~K.

Number densities in this region of the VALC model range from
n$>10^{16}$ down to $10^{13}$~cm$^{-3}$ and decrease roughly exponentially
with an scale height $\sim 130$~km. Thus the ionization parameter,
$U \sim F/n$ \citep[e.g.][]{Davidson.Netzer:79}, where $F$ is the
ionizing photon flux per unit area, is small for
even the strongest flares and X-ray photoionization of the
photospheric gas is inconsequential. This is in contrast to models of,
e.g., X-ray irradiated accretion disks
where the ionization state of the reflecting medium is governed by the
incident hard photon flux and must be computed self-consistently with
the emergent X-ray spectrum \citep[e.g.][]{Kallman.etal:04}.

The range of heights in the photosphere where the X-ray photons are
absorbed is very small compared with the stellar radius (of order
$\sim 100$-$1000$~km), and the variation in temperature and ionization
is also small over this altitude range (Figure~\ref{f:depth}).  We
therefore approximate the photosphere by a geometrically thin
but optically very thick shell of homogeneous
density.  We set the density to an arbitrarily high value such as to
obtain the same behaviour as for a plane-parallel model where the
photosphere is approximated by a cold semi-infinite slab with constant
density.  The appropriate atmospheric reference frame for describing
the absorption, scattering and fluorescent processes is the mass or
particle column depth and the calculations are independent of the
absolute value of the density.  Provided plane-parallel geometry
remains a good approximation, these calculations should therefore be
valid for stars with different surface gravities and atmospheric
pressures.

By assuming the photosphere to be cold, we make no distinction between
fluorescence processes in species of different charge states.  In
typical late-type stellar photospheres the dominant species are
neutral and once-ionized, for which the splitting between the
respective fluorescence lines is very small as are differences in
fluorescent yields \citep{House:69,Kallman.etal:04}.  We also
note that, as illustrated in Figure~\ref{f:cross}, in the energy range
of interest here there is essentially no difference in the
photoabsorption cross-sections for neutral or once-ionized species
except at low energies where the difference is dominated by He and
H. In reality, it is only the metals with fairly low first ionization
potentials that are significantly ionized in the upper photosphere and
lower chromosphere while He and H are almost entirely neutral.

\subsection{Monte Carlo (MC) Calculations of Fluorescent Processes} 
\label{s:monte}

We use a modified version of the MC photoionization and radiative
transfer code {\sc mocassin} \citep{Ercolano.etal:03,
Ercolano.etal:05}.  This code uses a stochastic approach to the
transfer of radiation which allows the construction of models of
arbitrary geometry where all components of the radiation field are
treated self-consistently.  The basic idea is to employ a discrete
description of the radiation field, whereby the simulation quanta are
monochromatic packets of radiation of constant energy,
$\varepsilon_0$, \citep{Abbott.Lucy:85}.  The individual processes of
scattering, absorption and re-emission of radiation can then be
simulated by sampling probability density functions based on the
medium opacities and emissivities, thus allowing the determination of
the trajectories of packets as they leave the illuminating source(s)
and diffuse outward.  These trajectories constitute the MC observable
of our experiment which can be related to the mean intensity of the
radiation field via the following MC estimator \citep{Lucy:99}:
\begin{equation}
4\pi\,J_{\nu}\,d\nu\,=\,\frac{\varepsilon_0}{\Delta t}\frac{1}{V}\sum_{d\nu}l,
\end{equation}
where $V$ is the volume of the current grid cell and the summation is
over all the fragments of trajectory, $l$, in $V$, for packets with
frequencies in the ($\nu$, $\nu + d\nu$) interval. $\Delta\,t$ is the
duration of the Monte Carlo experiment. The significance of this
estimator may be most easily understood by considering that 
the energy density of the radiation field in ($\nu$, $\nu + d\nu$) is 
$4\,\pi\,J_{\nu}\,d\nu/c$. At each given instant a packet contributes
energy $\varepsilon_0$ to the volume element containing it. The 
time-averaged energy content of a volume element is given by
the summation of the energy contribution of all packets crossing this 
volume element in the time interval $\Delta\,t$. Equation~1 follows
from this argument, since each packet
contributes $\varepsilon_0\,\delta\,t/\Delta\,t$ to the time
averaged energy content of a volume element and $\delta\,t$~=~$l/c$,
where $l$ is the segment of the trajectory of a packet contained in the
volume element. 

{\sc mocassin}'s atomic database 
is regularly updated and currently uses opacity data from
\citet{Verner.etal:93} and \citet{Verner.Yakovlev:95},  
energy levels, collision strengths and transition probabilities from
Version 5 of the CHIANTI database \citep[][and references
  therein]{Landi.etal:06} and the improved hydrogen and helium
free-bound continuous emission data recently published by
\citet{Ercolano.Storey:06}.  The \citet{Verner.Yakovlev:95}
photoabsorption cross-sections do no not include the resonance structure
that can be prominent near ionisation thresholds.  While this
structure can be important for some astrophysical applications
\citep[e.g.][]{Kallman.etal:04}, it is not significant for the
fluorescent efficiencies discussed here that depend primarily on 
broad-band spectral properties and neutral or once-ionised gas.

In the case of inner shell Fe\,K${\alpha}$ production, the fluorescent
luminosity, L(Fe\,K${\alpha}$), due to the absorption of an incident
energy packet of wavelength shorter than the Fe K edge
($h\nu~>~$~7.11~keV) is calculated on the fly. The event of absorption
of a high energy packet is immediately followed by the re-emission of
a number, $n$ of Fe\,K${\alpha}$ packets from the same event location. For
an incident packet of frequency $\nu$, carrying energy
$\varepsilon_0$ in the unit time $\Delta\,t$, the total Fe\,K${\alpha}$
emission is given by
\begin{equation}
L(Fe\,K{\alpha})~=~n
L'(Fe\,K{\alpha})~=~n\,\frac{\varepsilon_{0}}{\Delta\,t}
\frac{1}{h\nu}\,
\frac{\sum_i\kappa_{\nu}^{Fe^i}Y_{Fe\,K{\alpha}}^i\varepsilon_{Fe\,K{\alpha}}^i}
{\kappa_{\nu}^{gas}}R_{\alpha}
\end{equation}
where $\kappa_{\nu}^{Fe^i}$ and $Y_{Fe\,K{\alpha}}^i$ are the
absorption opacity and the Fe\,K${\alpha}$ yield of $i$-times ionised
iron, $\varepsilon_{Fe\,K{\alpha}}^i$ is the energy of the K$\alpha$ line
of i-times ionised iron ($\sim$6.4\,keV) and the summation is over all
abundant ionisation stages,
$\kappa_{\nu}^{gas}$ is the absorption opacity due to all other
abundant species and $R_{\alpha}$ is the branching ratio between
$K{\alpha}$ and $K{\beta}$ fluorescence (0.882:0.118,
\citealt{Bambynek.etal:72}). We use a value of 0.34 for
the fluorescence yields of neutral and once-ionised iron
\citep{Bambynek.etal:72, Krause:79, Kallman.etal:04}. 

The $n$ Fe\,K${\alpha}$ energy packets are emitted in random
directions as the fluorescent emission process can be assumed to be
isotropic. To ensure conservation, energy per unit time
$(\varepsilon_0/\Delta\,t~-~L(Fe\,K{\alpha}))$ must also be immediately
re-emitted from the event location, at a frequency determined by
sampling the local emissivity of the medium and transferred through the
grid as described by \citet{Ercolano.etal:03}.

The fates of the newly emitted Fe\,K${\alpha}$ packets are determined
by the absorption and Compton opacities encountered along their
diffusion paths. Fe\,K${\alpha}$ packets that are absorbed will be
transformed into diffuse field packets or emission line packets at
another energy and
will not contribute to the emerging spectrum at
6.4\,keV. Fe\,K${\alpha}$ packets that are Compton scattered will
experience a change of direction and a shift in frequency (by
approximately $(h \nu)^2$/$m_e c^2$), their new frequency and direction
being determined using the Klein-Nishina formulae. These packets may
therefore finally emerge and contribute to the low energy shoulder of
the Fe~K${\alpha}$ feature, or may undergo further scatterings and/or
absorptions. 

The emergent integrated and the direction-dependent spectral energy
distributions (SEDs) are finally calculated using all escaped energy
packets, including contributions from Compton reflected packets and
fluorescence line packets. Our method also allows us to easily
separate the various components of the emergent SED.

Since the study presented here is similar to the earlier
B79 work, we summarise in brief the differences
between these studies:

\begin{enumerate} 

\item B79 was aimed at understanding Fe~K$\alpha$ fluorescence in
solar flares; the possible utility of fluorescence for estimating the
photospheric Fe abundance (which was of course more uncertain that at
present) was emphasised.  The calculations presented here extend the
parameter space so as to cover conditions found in large stellar
flares and stars completely covered in X-ray emitting active regions;
our emphasis is on the use of fluorescence as a diagnostic of coronal
and flare geometry on distant unresolved stars.

\item Input parameters used here are more up-to-date in several
respects.  Input spectra have been computed using isothermal coronal
plasma models whereas B79 used an analytical approximation to the
bremsstrahlung continuum.  B79 also adopted a simple form for
absorption cross-sections of Fe and the photospheric gas.  Here,
photoabsorption cross-sections are computed self-consistently
according to the chemical composition and ionisation state of the gas
using analytical fits to Opacity Project data
\citep{Verner.Yakovlev:95}.  These abundance-weighted cross-sections
are about 40\%\ higher than that used by B79 for the gas with Fe
excluded; the B79 cross-section for neutral Fe K-shell absorption is
about 20\%\ larger than that of \citet{Verner.Yakovlev:95} at
threshold, but very similar at energies above 8~keV.  As we
shall see below, these differences have little influence on the
resulting fluorescence efficiencies, largely because the Fe
cross-section dominates and differences here are small.


\item B79 assumed that once a Fe~K$\alpha$ line photon is Compton
scattered it is ``lost'' due to the shift in energy. Here, we follow
all photons until they emerge from the photosphere and therefore also
resolve the Compton shoulders comprising scattered line photons.


\end{enumerate}

\subsection{Benchmark Calculations of Iron K$\alpha$ in solar flares}
\label{s:bench}

In order to establish the robustness of our new code, we first examine
the Fe\,K${\alpha}$ problem for point-like solar flares using the full
3D geometry of our models and fully self-consistent radiative
transfer, treating photoabsorption contributions from all abundant
atoms and ions, and (Compton) scattering by electrons.  As a
benchmark, we attempt to reproduce the earlier results of B79.


The photon density spectrum of the incident X-rays, $L(\varepsilon)$,
is taken to follow the same bremsstrahlung power-law with energy,
$\varepsilon$, as defined in B79:
\begin{equation}
\label{e:baispe}
L(\varepsilon)~=~A~\varepsilon^{-1}~\exp(-\varepsilon/kT) \; 
\rm photon~s^{-1}keV^{-1} 
\end{equation}
We consider four X-ray temperatures (0.5, 1.0, 3.0, 5.0~keV) and two
flare heights, $h=0$ and $h=0.1~R_{\odot}$, which cover the full parameter
space investigated by B79. In the remainder of this paper we will
quote temperatures in degrees K, however in this section these are
given in keV, which  
are the units used by B79. For the sake of
comparison, we used the Fe abundance adopted by B79 of $n_{\rm
Fe}$/$n_{\rm H}~=~4\times10^{-5}$ by number; for other elements we
adopted the mixture of GS.  
In Table~\ref{t:baiht} we compare our
results for the iron K$\alpha$ fluorescence efficiency, $\Gamma$, to
those of B79 (values in brackets). $\Gamma$ is defined as the ratio
of the total luminosity of iron K$\alpha$ photons emitted from the
photosphere and $N_{7.11}/2$, where
$N_{7.11}=\int_{7.11}^{\infty}L(\varepsilon) \, dE$, with $L(\varepsilon)$
given by Equation~\ref{e:baispe}.

In this formulation the flux of Fe\,K${\alpha}$ photons received at
Earth is equal to  
\begin{equation}
F_{K\alpha} = \frac{\Gamma f(\theta) N_{7.11}}{4\pi\,D^2 } ,
\label{e:feka}
\end{equation}
where D is the distance from the source and $\theta$ the heliocentric
angle subtended by the flare and observer.

The agreement with the results of B79 is generally very good, and
significantly better at higher X-ray temperatures; discrepancies
amount to only 18\%\ for X-ray temperatures of 0.5~keV. We attribute
the bulk of the differences in fluorescent efficiencies to the use by
B79 of slightly different absorption cross-sections and an analytical
approximation for the gas opacities, while we calculate these
self-consistently according to the elemental abundances considered in
our models. The slight temperature dependence in the differences is
most likely due to the different slopes of the B79 functional form and
our calculations, which take into account all elements with Z$<$30.

\begin{figure}
\begin{center}
\begin{minipage}[t]{7.5cm}
\includegraphics[width=7.5cm]{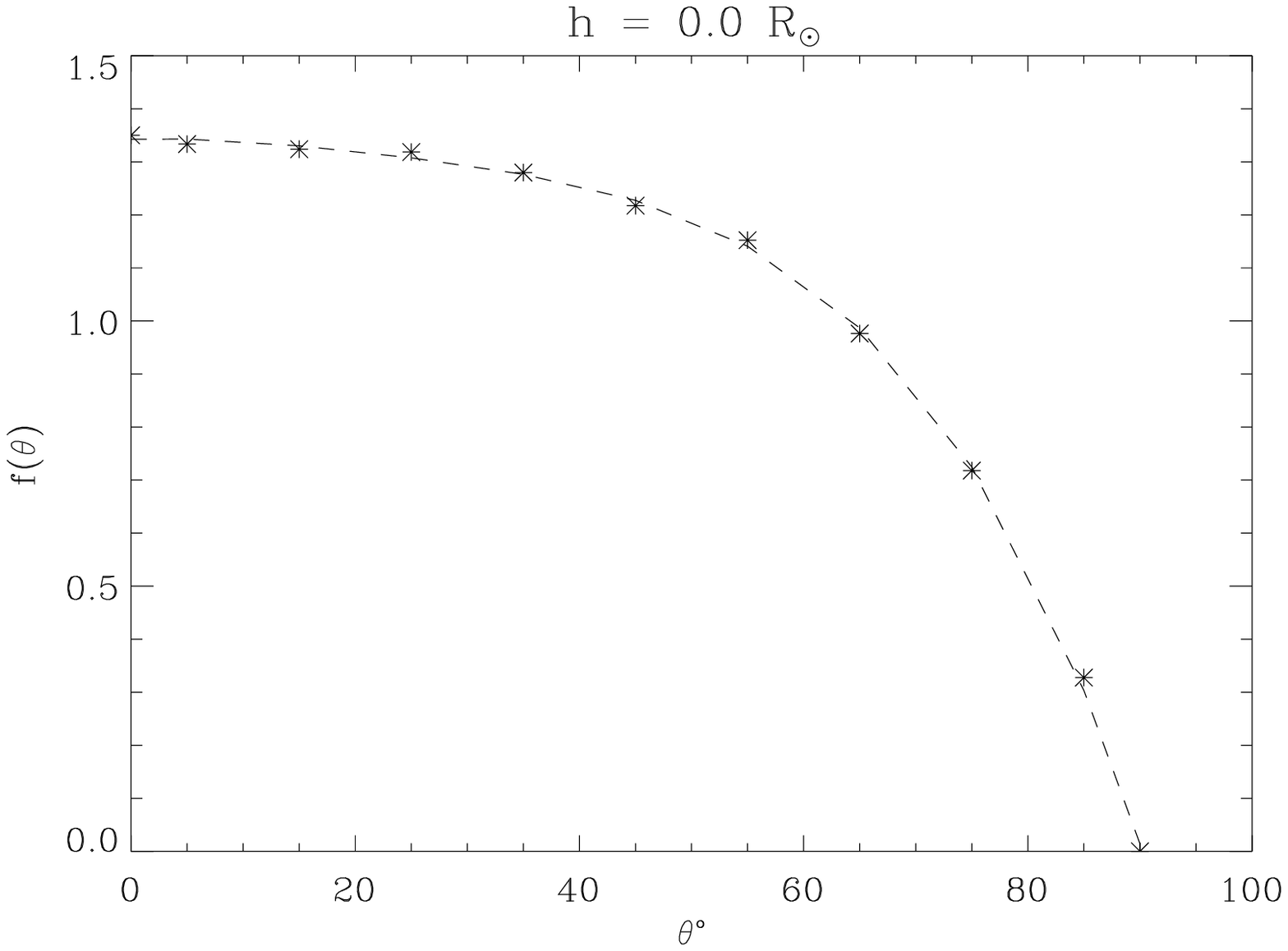}
\end{minipage}
\begin{minipage}[t]{7.5cm}
\includegraphics[width=7.5cm]{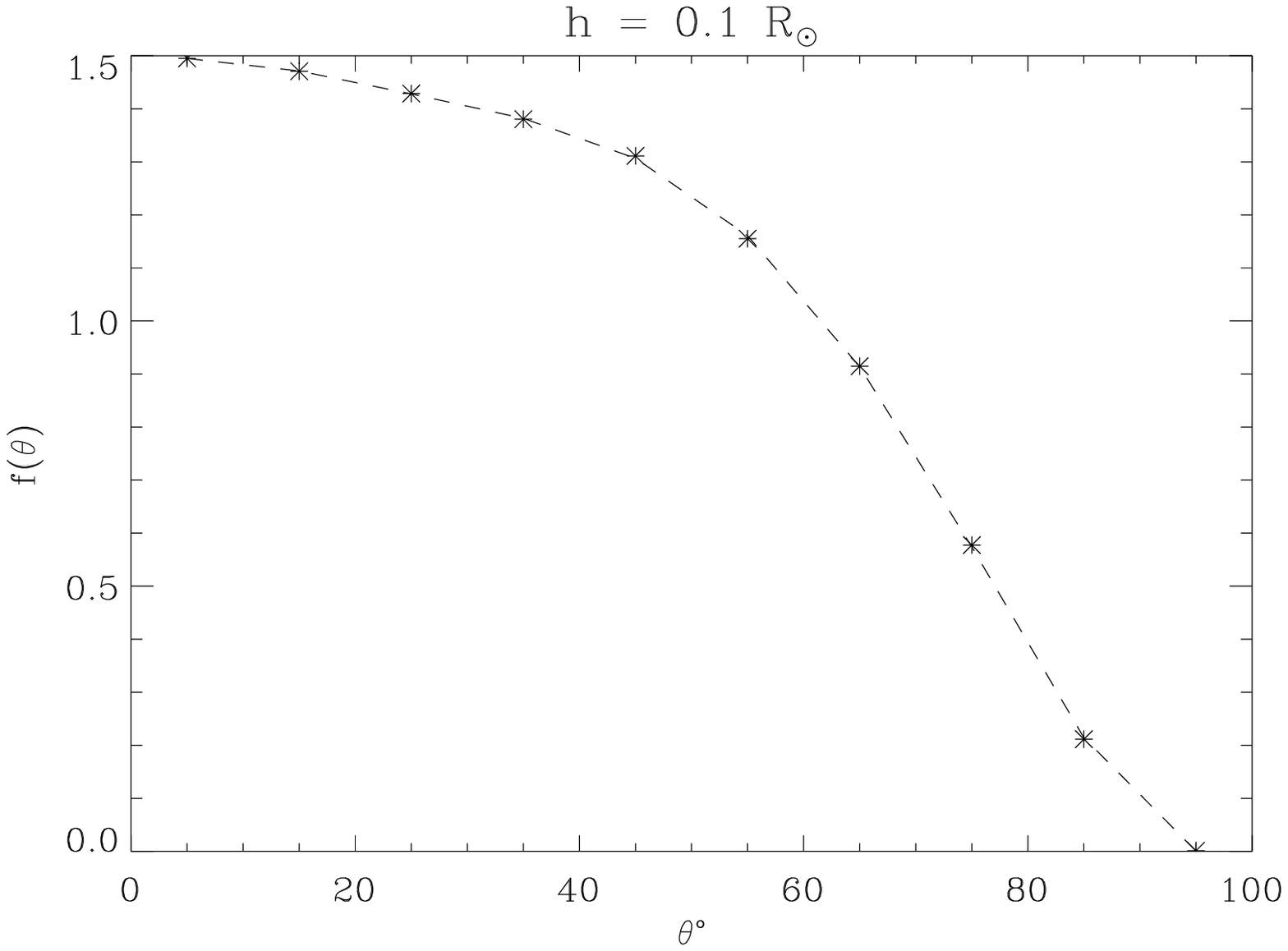}
\end{minipage}
\caption[]{The dependence of the Fe\,K${\alpha}$ fluorescence flux on
  the heliocentric angle, parametrized by the $f(\theta)$ factor, for two heights. 
These results are directly comparable to the top left and bottom right panels of Figure~3 of B79.}
\label{f:ftheta}
\end{center}
\end{figure}

Our 3D models allow us to obtain the spectral energy distribution
emerging from arbitrary viewing angles. In Figure~\ref{f:ftheta} we
show the variation of $f(\theta)$ for ten heliocentric angles from
$5^\circ$ to $95^\circ$ for two flare heights ($h=0$ and
$h=0.1R_{\odot}$). These plots are directly comparable with the
top-left and bottom-right panels of Figure~3 of B79 and show good
agreement with his results.

\begin{figure}
\epsscale{1.0}
\plotone{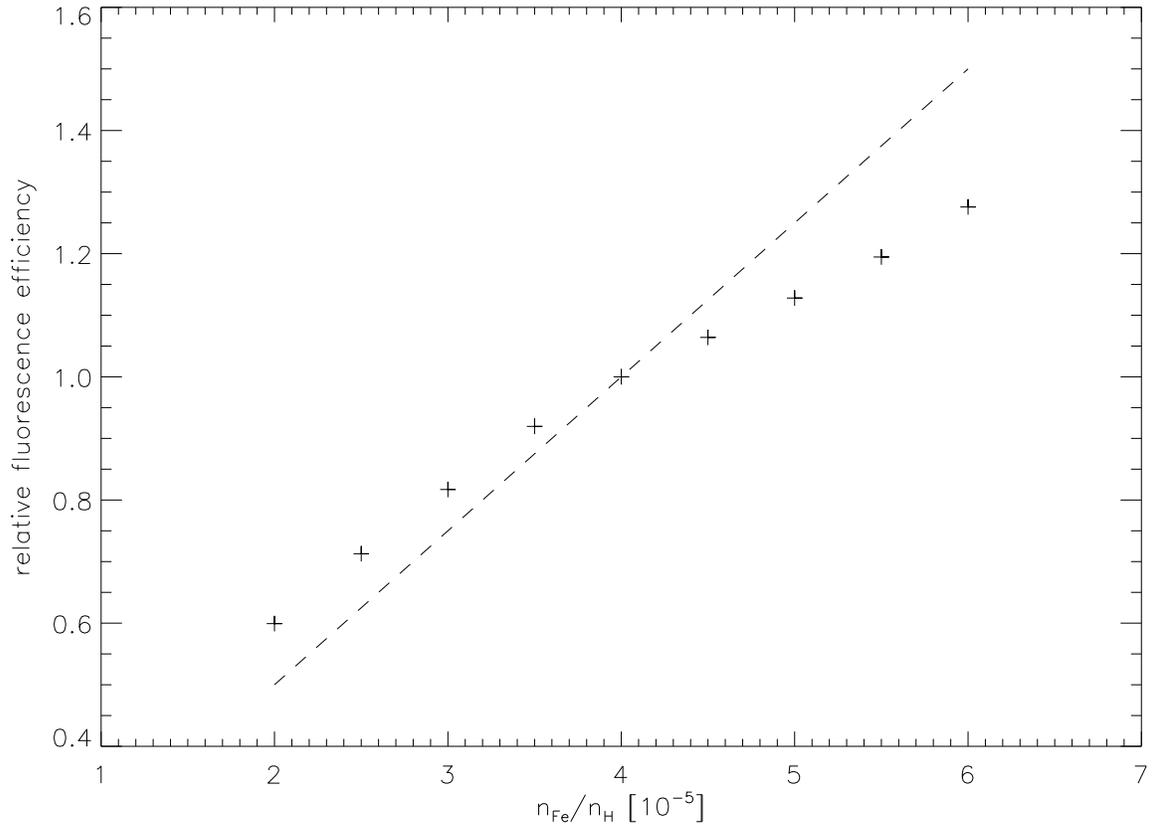}
\caption[]{Dependency of Fe\,K${\alpha}$ efficiency on 
the abundance of iron for the B79 benchmark case. The proportional relation is 
indicated by the dashed line. This figure is directly comparable to Figure~4 of B79.}
\label{f:abun}
\end{figure}

Finally, in Figure~\ref{f:abun} we show the variation of $\Gamma$ as
a function of iron abundance.  Again, our results are in good
agreement with those of B79 ({\it cf} his Figure~4) and show that the
relation of the Fe\,K${\alpha}$ fluorescence efficiency on the
photospheric iron abundance is weaker than a proportional relation, as
expected.  We discuss the Fe dependence of the Fe~K$\alpha$ flux in
more detail in \S\ref{s:fesens} below.

\section{New Fe\,K${\alpha}$ Diagnostic Calculations}
\label{s:newcalcs}

Having in the previous section established the robustness of our methods, 
we present the results from a new grid of model calculations designed to
provide useful diagnostics for Fe\,K${\alpha}$ detections for a wider range
of stellar environments.  In particular, stellar coronae cover a much
wider range of temperatures and Fe abundance than the solar case and
investigation of fluorescence over a wider range of parameters will be
necessary for correct interpretation of fluorescent lines from stars.

We investigate two geometrical configurations: (i) a single flare and
(ii) an active corona that can be approximated by a spherical shell
illuminating a photosphere from particular scale height $h$.  While
coronae are of course not expected to conform to a shell-like
geometry, this case should approximate quite well the case in which a
star is quasi-uniformly covered in active regions.  Coronal
fluorescing spectra were adopted from a grid of isothermal models
computed using emissivities from the CHIANTI compilation of atomic
data \citep[][and references therein]{Landi.etal:06}, together with
ion populations from \citet{Mazzotta.etal:98}, as implemented in
the PINTofALE IDL software suit \citep{Kashyap.Drake:00}.  We adopted
the chemical composition of GS for all calculations except where
noted.  While there will 
be some contribution from transitions in He-like and H-like Ni and the
higher Lyman series of He-like and H-like Fe to the ionising photon
flux $N_{7.11}$, this is small compared with the integrated continuum
contribution; consequently the metallicity sensitivity for the
ionising spectrum is generally negligible for realistic ranges of
metallicity.

\subsection{A Single Flare Illuminating a Photosphere}
\label{s:flare}

In the case of a single flare illuminating a photosphere from a given
height, the heliocentric angle, $\theta$, has a dramatic effect on the
Fe\,K${\alpha}$ flux detected. The intrinsic Fe\,K${\alpha}$
efficiency, $\Gamma$, is, of course, independent of $\theta$, being a
function of the fluorescing spectrum, the flare height and the
relative iron abundance. Using GS solar abundances for all
elements with $Z\,<\,30$ we show in Table~\ref{t:feka1} the
Fe\,K${\alpha}$ efficiency, $\Gamma$, predicted for X-ray
temperatures in the range $T=6.4$--8.0~keV and flare
height in stellar radii in the range $h=0$--$5R_*$.

\begin{figure}
\epsscale{1.0}
\plotone{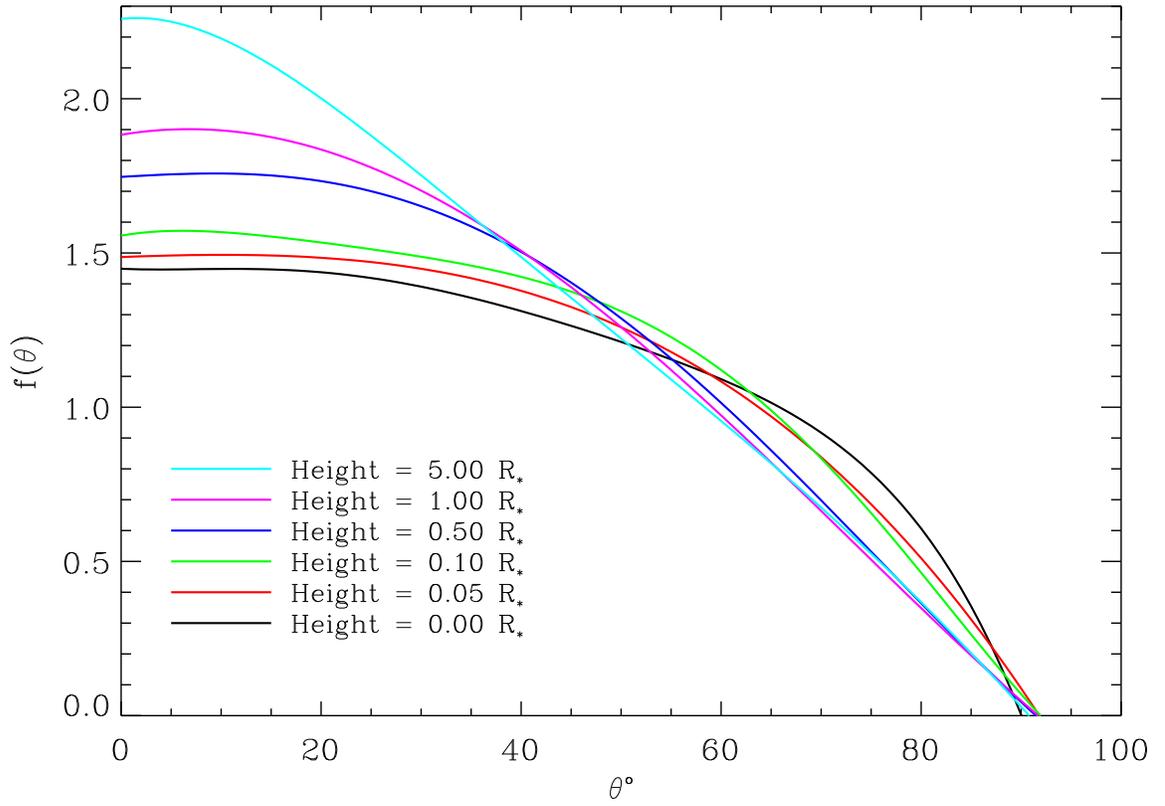}
\caption[]{Fits to the $f(\theta)$ factor relating the Fe\,K${\alpha}$
  fluorescence flux to the angle of observation obtained for six flare
  heights}
\label{f:fthetafits}
\end{figure}

In order to investigate the Fe\,K${\alpha}$ line strength as a
function of heliocentric angle, we calculated $f(\theta)$ for a number
of angles at each flare height.  For convenience we fit our results
using 4th or 5th order polynomials, such that
$f(\theta^o)~=~\sum_{i=0}^5\alpha_i\cdot\theta^i$.  The fit
coefficients are listed in Table~\ref{t:fthetafits} and the resulting
$f(\theta)$ functions are plotted in Figure~\ref{f:fthetafits}. Our
fits are valid between heliocentric angles 0$^{\circ}$ and
100$^{\circ}$; the maximum errors between our data and the fits at
each flare height are also listed as percentages in
Table~\ref{t:fthetafits} and are less than 2.4\%. 

These functions illuminate interesting characteristics of the
fluorescence problem for different incidence and exit angles.
Firstly, we point out that for finite flare heights $h$, $f(\theta)$
is non-zero for angles $\theta > 90^\circ$.  This is simply due to the
surface illumination from flares beyond the limb extending onto the
visible hemisphere: the angle at which $f(\theta)$ reaches 0 is given
geometrically by $\theta_0=90^\circ+arccos(R_\star/(R_\star +h))$. We
note, however, that we did not explore the very low-response tail of
$f(\theta)$ for heliocentric angles greater than $100^{\circ}$, as the
Fe~K$\alpha$ flux, although finite for some flare heights, is too
small to be detectable even for very strong flares.  Furthermore,
$f(\theta)$ falls below the statistical variance of our models at
these large angles.  

The general decrease of $f(\theta)$ with
increasing $\theta$ present in all curves is largely a result of the
$1/cos(\theta)$ increase in path length required for K$\alpha$ photons
to escape.  The most conspicuous characteristic of $f(\theta)$
behaviour with different heights is the steepening of the decline from
smaller to larger angles with increasing flare height.  This arises
because of the range of incidence angles of flare photons on the
photosphere.  As $h\rightarrow \infty$, incidence angles relative to
the photospheric normal become smaller and the average penetration
depth before interaction by either absorption or scattering events is
larger.  Consequently, path lengths increase and Fe~K$\alpha$ escape
probabilities decrease.  Folded in with this is the decreasing
fraction of photons incident on the visible hemisphere of the star
with increasing $\theta$.

\subsection{A Coronal Shell Illuminating a Photosphere}
\label{s:shell}

The calculations for $f(\theta)$ and $\Gamma$ can be readily adapted
from the point source case to one in which the corona is approximated
by a spherical shell of emission, as might be the case for stars
significantly more active than the Sun.  This can be seen by taking
the shell case to be a large ensemble of point sources distributed
uniformly about the star for which the observed fluorescent flux is
then simply the solid angle average of the point source case,
\begin{equation}
F_{K\alpha}=\frac{\Gamma N_{7.11}}{4\pi D^2} \frac{1}{4\pi}
\int f(\theta) \; d\Omega
\label{e:shell}
\end{equation}
Since, following from the definition of $\Gamma$ and our Equation~4,
$\int f(\theta) \; d\Omega= 2\pi$, the flux is 
\begin{equation}
F_{K\alpha}=\frac{\Gamma N_{7.11}}{8\pi D^2}
\end{equation}
By tailoring the solid angle integral in Equation~\ref{e:shell} the
fluorescent flux for an arbitrary coverage pattern of active regions
viewed from any angle can be derived from the tabulated values of
$f(\theta)$ and $\Gamma$. 

\subsection{Sensitivity to Ionising X-ray Temperature}
\label{s:tsens}

\begin{figure}
\plotone{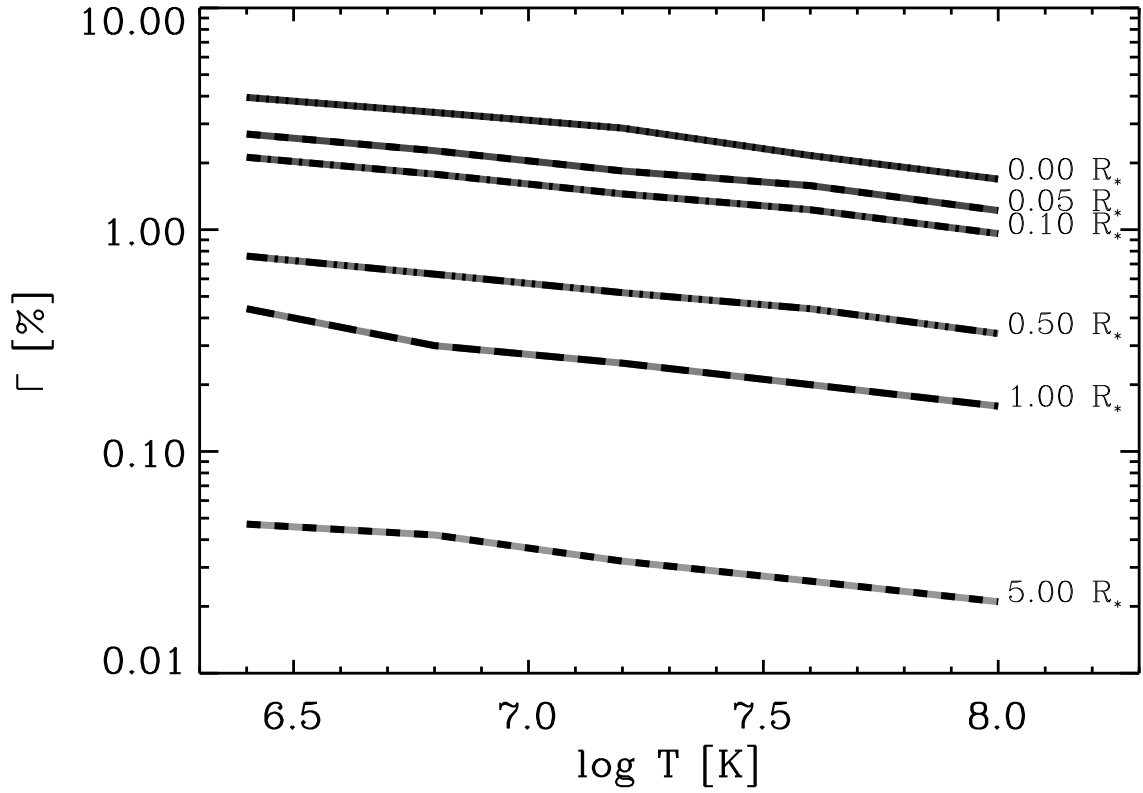}
\caption{The fluorescence efficiency, $\Gamma$, as a function of
  the ionising X-ray spectrum temperature for different flare heights, 
computed for the GS solar abundance mixture.
}
\label{f:vst}
\end{figure}

The fluorescence efficiency, $\Gamma$, is illustrated as a function of
the isothermal plasma temperature adopted for the ionising coronal
spectrum in Figure~\ref{f:vst}.  We find $\Gamma$ varies very slowly
with temperature, decreasing by only a factor of $\sim 2$ over a
factor of 30 or more in temperature.  This gradual decrease is
predominantly a result of the competition between Fe inner-shell
absorption and Compton scattering.  As temperatures increase, incident
photons are present at higher and higher energies.  Since the Fe
cross-section declines with increasing photon energy while the Compton
cross-section remains constant, higher energy photons are more liable
to be scattered out of the photosphere before being destroyed by
photoabsorption.

\subsection{Sensitivity to Photospheric Fe
  Abundance and Metallicity} 
\label{s:fesens}

In the optically-thick fluorescence case such as characterizes the
photospheric fluorescence problem, inner-shell photoabsorption by Fe
atoms and ions competes with the background opacity from other
elements.  Observed Fe~K$\alpha$ line fluxes and equivalent widths are
then expected to vary predominantly according to the {\em relative} Fe
abundance rather than just the photospheric metallicity.  Note that
the optically-thick situation differs to the optically-thin formalism
discussed by \citet{Liedahl:99} and referred to in the discussion of
stellar and protostellar Fe~K$\alpha$ lines
\citep[e.g.][]{Favata.etal:05,Tsujimoto.etal:05,Osten.etal:07}: in the
optically-thin case the fluorescent Fe line strength is primarily a
function of the absorbing Fe column.

\begin{figure}
\plotone{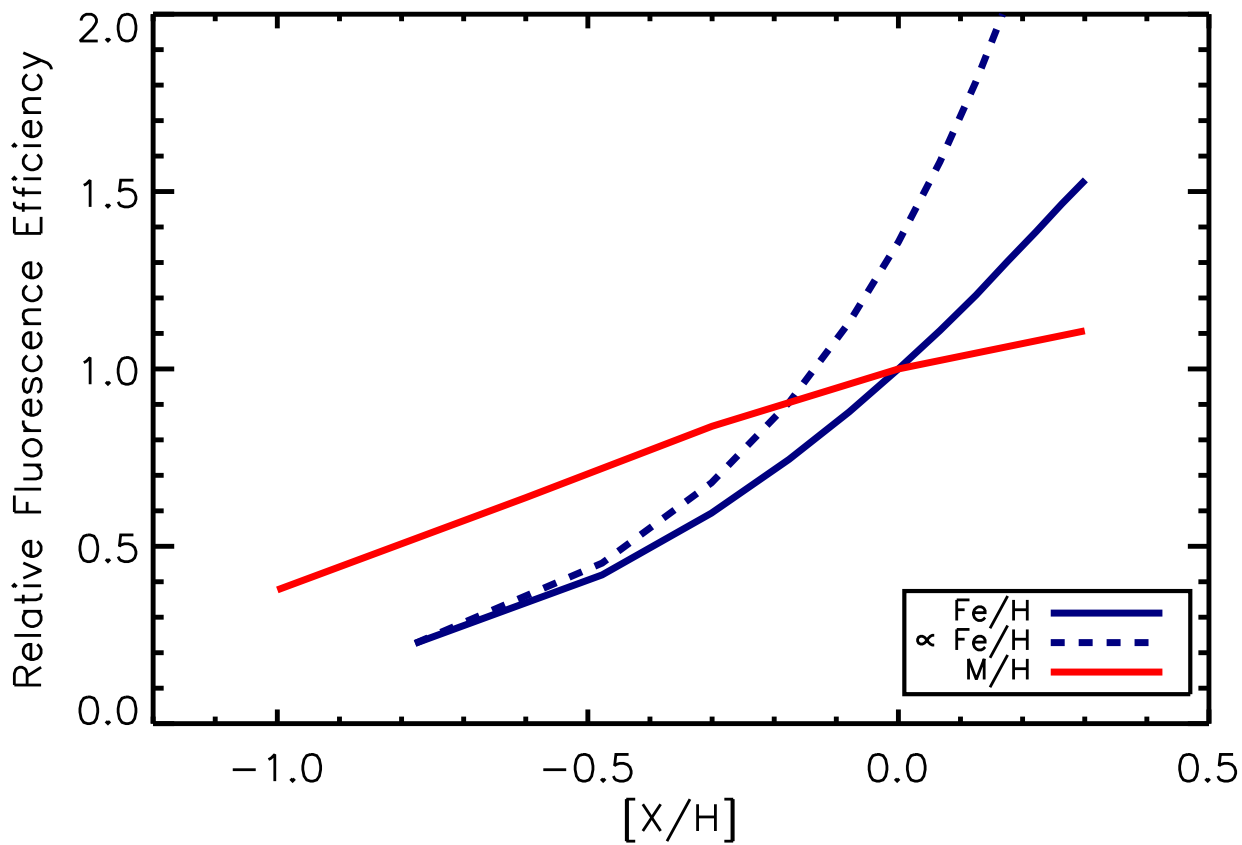}
\caption{The fluorescence efficiency, $\Gamma$, as a function of
  logarithmic photospheric metallicity [M/H] and Fe abundance [Fe/H],
  normalised to its value for the solar mixture of GS.  The
  efficiencies as a function of [Fe/H] were computed for GS
  values for the abundances of all other elements.  The dashed curve
  represents the proportionality relation expected from scaling the
  efficiency linearly with Fe abundance from the lower Fe abundance
  point.  Calculations where
  performed for an ionising X-ray spectrum with temperature of 10$^{7.2}$~K.
}
\label{f:vsfeabun}
\end{figure}

In extension to the benchmark calculations discussed in
\S\ref{s:bench}, we have computed the sensitivity of $F_{K\alpha}$ to
the Fe abundance for a model fluorescing spectrum with a temperature
of 10$^{7.2}$~K and Fe abundance in the range $-0.8 \leq
$[Fe/H]$\leq+0.3$ ($0.5\times 10^{-5} \leq n_{\rm Fe}/n_{\rm H} \leq
6\times 10^{-5}$).  This temperature was chosen to be representative of
a typical very active corona or moderate stellar flare.
Trends with [Fe/H] are expected to be only 
weakly sensitive to this adopted temperature, the fluorescence
efficiency, $\Gamma$, varying only very slowly with $T$
(Figure~\ref{f:vst}).  The results are illustrated in
Figure~\ref{f:vsfeabun} where we show $\Gamma$ normalised to its value
at the GS solar Fe abundance in comparison to the proportional
relation, analogous to Figure~\ref{f:abun} discussed in
\S\ref{s:bench} and Figure~4 of B79.

The fluorescence line strength vs Fe abundance was investigated in the
context of accretion disks by \citet{Matt.etal:97}, and our results
are comparable to Figures~1 and 2 from that work for the range of Fe
abundance in common.  As noted above, B79 showed that the relationship
between the relative fluorescence efficiency $\Gamma$ and the Fe
abundances is weaker than a proportional relation.  In fact, the
relation increasingly departs from proportionality with increasing Fe
abundance.  This behaviour arises because of the equipartition of
photoabsorption between the different species present in the
atmosphere.  At low Fe abundance when the background cross-section due
to elements other than Fe dominate, $F_{K\alpha}$ varies according to
$n_{\rm Fe}\sigma_{\rm FeK}/\sigma_{tot}$, where $\sigma_{\rm FeK}$
and $\sigma_{tot}$ are the Fe $n=1$ shell and total plasma absorption
cross-sections, respectively.  At very large Fe abundance, the
photoionization cross-section of Fe will begin to
dominate $\sigma_{tot}$; $\sigma_{\rm FeK}/\sigma_{tot}$ then tends to
asymptote to the constant ratio $\sigma_{\rm FeK}/(\sigma_{\rm
  FeK}+\sigma_{\rm FeL,M})$, where 
$\sigma_{\rm FeL,M}$ is the total Fe photoabsorption cross-section
from levels $n > 1$.  We do not consider such large Fe abundances
here; this asymptotic behaviour is nicely illustrated by
\citet{Matt.etal:97}.  The relationship is further complicated to some
extent by the
role of Compton scattering, but this becomes less relevant for higher Fe
abundances for which photoabsorption dominates Compton scattering out
to higher energies.

Also illustrated in Figure~\ref{f:vsfeabun} is the relative
fluorescence efficiency for different photospheric {\em metallicity}
(ie where all metal abundances are scaled together) for the range
$-1.0\leq $[M/H]$\leq +0.3$ relative to the GS solar mixture.  As
expected, the fluorescent efficiency is not so sensitive to [M/H]
because of the tendency of $F_{K\alpha}$ to follow the relations above
between the different opacity sources: adjusting the metallicity as a
whole results in a smaller change in $n_{\rm Fe}\sigma_{\rm
FeK}/\sigma_{tot}$ since $\sigma_{tot}$ is by definition in lockstep
with metallicity.  As metallicity decreases, Compton scattering and
the background H and He cross-sections become more important, the
combined effect of which is to weaken $F_{K\alpha}$ in
comparison to the higher metallicity case.  For very high
metallicities, these opacity sources are negligible and the
fluorescence efficiency asymptotes to the value given by 
$n_{\rm Fe}\sigma_{\rm FeK}/\sigma_{tot}$.

%
%
%
%

\section{Discussion and Applications}

The main scientific motivation for this work is to provide the
foundation to use Fe fluorescence as a quantitative diagnostic of
coronal and flare geometry.  There now exists a
handful of detections of fluorescent emission from stars.  Sensitivity
is currently limited to a large extent by the low spectral resolution of
available instruments and progress is expected to accelerate
dramatically with the future availability of X-ray calorimeters.

\begin{figure}
\plotone{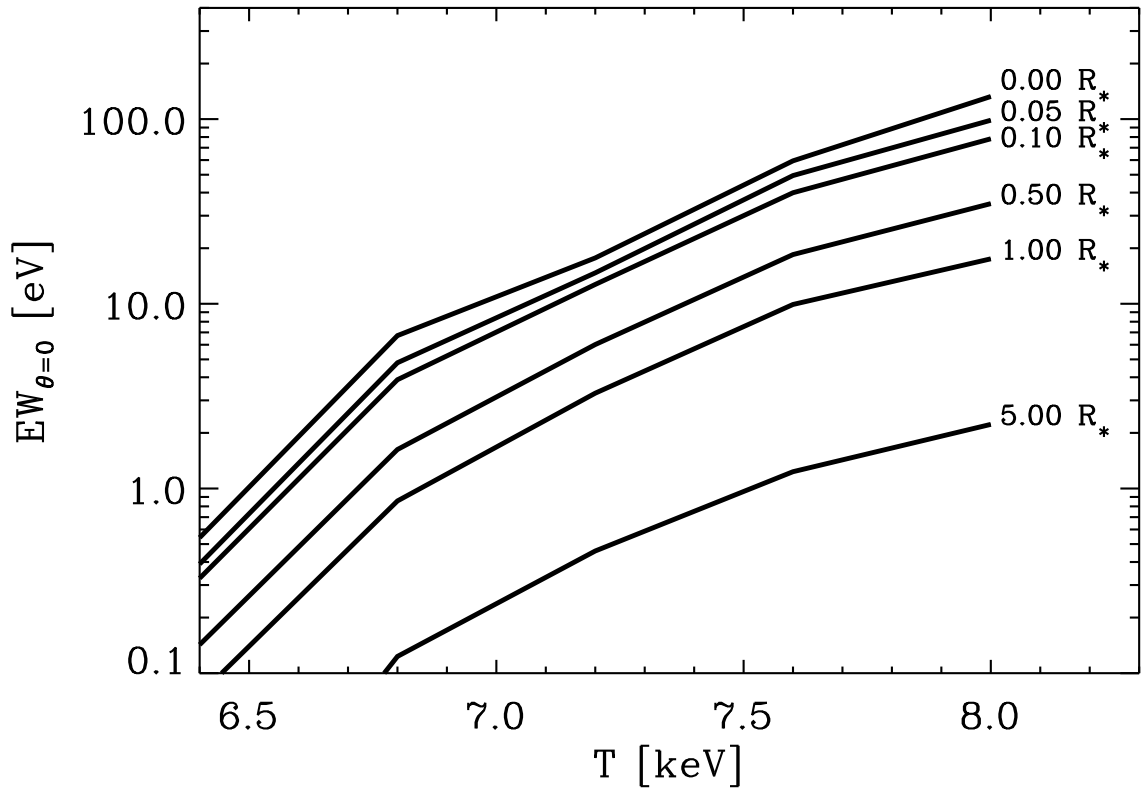}
\caption{The Fe~K$\alpha$ fluorescent line equivalent width at $\theta$~=~0 as 
a function of X-ray temperature and flare height, computed for the GS solar abundance mix.}
\label{f:ew}
\end{figure}

Since modern X-ray spectral analyses
based on low-resolution CCD pulse-height spectra tend to express line
strengths in terms of the line equivalent width, we have computed this
quantity for the case of $\theta=0$ and the ranges of heights and
X-ray temperatures investigated in \S\ref{s:newcalcs}.  The equivalent
width in this context refers to the fluorescent, processed line seen
on top of the continuum of the ionising coronal spectrum.
While the
$\theta=0$ case gives the most optimistic line strength, we note that
$f(\theta)$ is quite slowly varying for angles $\theta \la 45^\circ$
for the flare heights for which significant Fe\,K${\alpha}$ might be
observed.  The equivalent widths are illustrated in Figure~\ref{f:ew}
and listed in Table~\ref{t:fekaEW1}.

\subsection{Fluorescence from Pre-Main Sequence Stars}

The observability of the cold Fe~K$\alpha$ line is of course strongly
dependent on the quality of the X-ray spectrum obtained.  The most
extensive study of PMS Fe fluorescence to date is that
based on Chandra observations of the Orion Nebula Cluster by
\citet{Tsujimoto.etal:05}.  This study detected significant 6.4~keV
excesses attributable to Fe fluorescence for 7 out of 127 sources
found to have significant counts in the 6-9~keV band.  Equivalent
widths were in the range 110-270~eV at plasma temperatures of $\sim
3$-10~keV.  There is clearly a strong selection effect here and these
fluorescent line strengths likely represent the upper end of the
distribution.  

Our calculations for a flare at scale height $h=0 R_\star$ are also
appropriate for a flare occurring above an infinite plane, such as
might approximate a disk-encircled PMS star.  
As in the photospheric case, the fluorescence problem can be treated 
orthogonally from the ionisation structure of the disk, which is not 
greatly altered from its very largely neutral overall state 
by X-rays from a typical flare.  Any small degree of X-ray
photoionisation will also not affect Fe~K$\alpha$ line strengths
because fluorescence yields are essentially
invariant for lower Fe ions.
Our calculations indicate that attaining
equivalent widths much in excess of 100~eV is not straightforward for
such a simple geometry for the plasma temperatures observed in the
fluorescing X-ray spectra.  This finding is in agreement with earlier
calculations by \citet{Matt.etal:91} and \citet{George.Fabian:91}, who find
equivalent widths of $\sim 150$~eV for a flat disk illuminated by
X-rays with power-law photon spectral energy distributions.

There are at least four ways in which equivalent widths might be
elevated above the values we find: (1) super-solar Fe abundance in the
disk material, possibly arising as a result of an elevated dust-to-gas
ratio; (2) disk flaring, resulting in a solid angle coverage $> 2\pi$;
(3) line-of-sight obscuration of the central flaring source (but not
fluorescent line photons) by optically thick structures such as the
star itself (ie the flare occurring on the far hemisphere); and (4)
fluorescence contributions from ionisation by non-thermal electrons.

By analogy with the solar case, in which excitation by non-thermal
electrons is usually negligible \citep{Parmar.etal:84,Emslie.etal:86}
and is much more difficult on energetic grounds, we consider (4) the
{\em least} plausible of these.  \citet{Ballantyne.Fabian:03} have
also shown in the accretion disk context that Fe~K production by
non-thermal electron bombardment requires 2-4 orders of magnitude
greater energy dissipation in the electron beam than is required for
an X-ray photoionization source.

Disk flaring can give rise to increased line strengths by factors $<
2$ simply from consideration of the increased solid angle coverage
possible compared with an infinite flat disk.  It is also difficult to
envisage enhanced Fe abundances in the disk being able to elevate line
strengths by more than a factor of a few.  Of some interest, then, is
the observation of an Fe~K$\alpha$ line equivalent width of
$\approx1400$~m\AA\ during the rise phase of a flare on the PMS Orion
nebula star forming region object V~1486~Ori by
\citet{Czesla.Schmitt:07}.  Such an enhancement over an infinite disk
value of $\sim 150$~m\AA , even with a large degree of disk flaring,
would still require extreme enhancements of the disk Fe abundance by
an order of magnitude or more \citep[e.g. \S\ref{s:fesens}
  and][]{Matt.etal:97} were the line due to photoionisation by the
{\em directly observed} continuum.  We point out, however, that
fluorescent line photons from a PMS disk can still be observed when
the X-ray flaring source is located behind the star and obscured from
the line-of-sight.  In the case of the V~1486~Ori flare, the large
observed Fe~K$\alpha$ equivalent width is simply and plausibly
associated with a partially obscured flare whose rise phase was not
fully observed directly owing to line-of-sight obscuration by the star
itself.  Such obscured flares will inevitably be the cause of some
fraction of observed Fe~K$\alpha$ lines from PMS stellar disks.  This
explanation is also more consistent than one relying on preferential
disk geometry and Fe abundance with the non-detection of Fe~K$\alpha$
from a second less extreme flare whose impulsive phase instead appears
to have been quite visible.


\subsection{Fluorescence from stellar photospheres}


The strong flare in II~Peg observed by {\it Swift} and analysed by
\citet{Osten.etal:07} presents another interesting case.  Photospheric
Fe~K$\alpha$ was clearly detected throughout the event.  Equivalent
widths for different times in the flare ranged from 18 to 61~eV, with
uncertainties in the 20-45\%\ range.  The authors favoured a
collisional excitation mechanism for the line, arguing that
fluorescence would be unlikely to produce an observable feature.  This
assessment employed a simple analytical formula applicable to
optically-thin cases in which only a minor fraction of the incident
X-ray flux is subject to photoabsorption or scattering \citep[see
  e.g.][]{Liedahl:99,Krolik.Kallman:87}.  \citet{Osten.etal:07}
correctly noted that the path length required to obtain the observed
equivalent widths under such an approximation was similar to the
$\tau=1$ Compton scattering depth, but discounted fluorescence as a
possibility on these grounds.  Other than the inapplicability of the
optically-thin formula for the photospheric fluorescence case, one
reason such an argument is overly pessimistic is that incidence angles
on the photosphere range from $\sim 0$--$90^\circ$ for small scale
heights and path lengths for escape are less than penetration depths
by the factor of the inverse cosine of these angles.  We defer a more
detailed treatment of the event to future work, but note here that
equivalent widths of 50~eV are achieved for flare heights up to $\sim
0.2 R_\star$ or so for the $10^8$~K model in our grid, a temperature
similar to the average of the values found for the flare by
\citet{Osten.etal:07}.  While collisional ionisation cannot be ruled
out observationally as the source of the observed Fe fluorescence, it
is not a requirement.

\section{Summary}

We have investigated the production of Fe~K$\alpha$ fluorescence lines
by irradiation from coronal and flare emission using a 3-dimensional
Monte Carlo radiative transfer approach including Compton redistribution.  
The results are presented in the form of convenient tables describing
the fluorescence efficiency as a function of the flare height, the
temperature of the ionising X-ray spectrum, and the viewing angle.
We have also illustrated how the fluorescent efficiencies scale with
the photospheric metallicity and the Fe abundance.   The results
should be of use for interpreting observations of Fe~K$\alpha$ lines
seen from stars.  

Our computed Fe~K$\alpha$ equivalent widths for irradiation from a
height of zero above the photosphere correspond to the case of flares
above a disk of infinite extent and are relevant to observations of
fluorescence from PMS stars.  Observed equivalent widths
tend to be slightly larger than our computed ones for a plasma of solar
composition and are an indication of flaring disk geometries or
super-solar Fe abundances.  For one extreme case recently observed on
V~1486~Ori, we propose that the very large equivalent width observed
($\approx 1400$~m\AA) arose from X-ray fluorescence by a flare
partially obscured from the line-of-sight by the stellar limb.

The FeK$\alpha$ equivalent width reported for a large flare on II~Peg
is consistent with our computed values for a flare scale height of a
few tenths of a stellar radius, and a collisional excitation mechanism
is not a requirement.

\section*{Acknowledgments}

JJD was supported by the Chandra X-ray Center NASA contract NAS8-39073
during the course of this research and thanks the Director,
H.~Tananbaum, for continuing support and encouragement.  BE was
supported by {\it Chandra} grants GO6-7008X and GO6-7098X.  We thank
the NASA AISRP for providing financial assistance for the development
of the PINTofALE package and Dr.~Dima Verner for making available his
{\sc fortran} routine for calculating photoionization cross-sections.
Finally, we thnk the anonymous referee for comments and corrections
that improved the manuscript.


\newpage


\section*{Tables}

\begin{table}
\begin{center}
\caption{Fluorescence efficiencies, $\Gamma$ $[$\%$]$, for 
B79 benchmark models. B79
values are given in parentheses. The iron abundance $n_{\rm Fe}$~=~4\,$\cdot$10$^{-5}\, n_{\rm H}$ was used. }
\label{t:baiht}
\begin{tabular}{lcccc}
\hline
                      &  \multicolumn{4}{c}{X-ray temperature (keV)}\\
Height ($R_{\odot}$)  & 0.5 & 1.0 & 3.0 & 5.0                        \\
\hline\noalign{\smallskip}
0.                    &  4.06 (3.44) & 3.70 (3.26) & 2.93 (2.74) & 2.47 (2.44) \\
0.1                   &  1.99 (1.78) & 1.80 (1.65) & 1.39 (1.37) & 1.16 (1.22) \\
\hline\noalign{\smallskip}
\end{tabular}
\end{center}
\end{table}

\begin{table}
\begin{center}
\caption{Fluorescence efficiencies, $\Gamma$ $[$\%$]$, for a range of
  X-ray temperatures, $T_X$, and flare heights. \citet{Grevesse.Sauval:98}
  (GS) solar abundances were adopted for all elements. }
\label{t:feka1}
\begin{tabular}{lccccc}
\hline
                      &  \multicolumn{5}{c}{X-ray temperature, Log($T_X$) [K]}\\
Height ($R_{*}$)      & 6.4          & 6.8         & 7.2         & 7.6        & 8.0 \\
\hline\noalign{\smallskip}
0.00                  &  3.95        &  3.38     &  2.87     &  2.16    &  1.69  \\
0.05                  &  2.70        &  2.27     &  1.84     &  1.58    &  1.22  \\
0.10                  &  2.12        &  1.78     &  1.45     &  1.23    &  0.96  \\
0.50                  &  0.76        &  0.63     &  0.52     &  0.44    &  0.34  \\
1.00                  &  0.44        &  0.30     &  0.25     &  0.20    &  0.16  \\
5.00                  &  0.047       &  0.042    &  0.032    &  0.026   &  0.021 \\
\hline\noalign{\smallskip}
\end{tabular}
\end{center}
\end{table}

\begin{table*}
\begin{center}
\caption{Polynomial fit coefficients to heliocentric angle
  functions. The maximum error between our fit and the data is given
  in the last common as a percentage value.}
\label{t:fthetafits}
\begin{tabular}{lccccccc}
\hline
Height ($R_{*}$)      & $\alpha_0$ & $\alpha_1$ & $\alpha_2$ &
$\alpha_3$   & $\alpha_4$ & $\alpha_5$  & $E_{max} [\%]$\\
\hline\noalign{\smallskip}
0.00                  &  1.45       &-1.53E-3   &  2.92E-4
&-1.73E-5  &  2.88E-7 & -1.68E-9& 0.6  \\
0.05                  &  1.49       & 1.26E-3   & -3.97E-5    &-1.40E-6  & -2.54E-9 &  0.     & 1.6 \\
0.10                  &  1.56       & 5.78E-3   & -6.37E-4    & 2.02E-5  & -3.07E-7 &  1.45E-9& 1.2 \\
0.50                  &  1.75       & 1.90E-3   & -2.17E-5    &-6.37E-6  &  5.09E-8 &  6.63E-11& 2.3\\
1.00                  &  1.88       & 5.36E-3   & -3.99E-4    &3.46E-7  &  1.01E-8 &  0.       & 0.8\\
5.00                  &  2.25       & 9.00E-3   & -1.75E-3    & 4.19E-5  & -4.69E-7 &  1.94E-9& 2.3\\
\hline\noalign{\smallskip}
\end{tabular}
\end{center}
\end{table*}

\begin{table}
\begin{center}
\caption{FeK$\alpha$ line equivalent widths (EW) at $\theta~=~0$ for a range of
  X-ray temperatures, $T_X$, and flare heights. \citet{Grevesse.Sauval:98}
  (GS) solar abundances were adopted for all elements. }
\label{t:fekaEW1}
\begin{tabular}{lccccc}
\hline
                      &  \multicolumn{5}{c}{X-ray temperature, Log($T_X$) [K]}\\
Height ($R_{*}$)      & 6.4          & 6.8         & 7.2         & 7.6        & 8.0 \\
\hline\noalign{\smallskip}
0.00                  &    0.54   &  6.73  &   17.73  &   59.45 &   132.57  \\
0.05                  &    0.39   &  4.79  &   14.71  &   49.49 &    98.66  \\
0.10                  &    0.33   &  3.88  &   12.73  &   39.95 &    78.23  \\
0.50                  &    0.14   &  1.63  &   6.02   &   18.47 &    34.87  \\
1.00                  &    0.076  &  0.86  &   3.28   &   9.92  &    17.52  \\
5.00                  &    0.0054 &  0.12  &   0.46   &   1.23  &     2.22  \\
\hline\noalign{\smallskip}
\end{tabular}
\end{center}
\end{table}

\clearpage

\section*{Figures}

\end{document}